\documentstyle[11pt,aaspp4,flushrt]{article}
\def\etal{{\it et~al.~}}
\def\eg{{\it e.g.~}}

\def\Sec{\rlap{$^{\prime\prime}$}.\hbox to 2pt{}}
\def\Min{\rlap{$^\prime$}.\hbox to 1pt{}}
\def\vmi{$V-I~$}

\begin{document}

\title{{\it Hubble Space Telescope} Observations of M32:
The Color-Magnitude Diagram\footnote[1]{Based on observations
with the NASA/ESA Hubble Space Telescope, obtained at the Space
Telescope Science Institute, which is operated by AURA, Inc., under
NASA contract NAS 5-26555.}}

\author{Carl J. Grillmair}
\affil{Jet Propulsion Laboratory, California Institute of Technology,
4800 Oak Grove Drive, Pasadena, California 91109-8099}

\author{Tod R. Lauer} 
\affil{Kitt Peak National Observatory, National Optical Astronomy
Observatories,\footnote[2]{Operated
by AURA, Inc., under cooperative agreement with the National Science
Foundation} P.O. Box 26732, Tucson, Arizona 85726}

\author{Guy Worthey\altaffilmark{3}} 
\affil{Department of Astronomy, University of Michigan, 
Ann Arbor, Michigan 48109-1090}
\altaffiltext{3}{Hubble Fellow}

\author{S. M. Faber}
\affil{UCO/Lick Observatory, Board of Studies in Astronomy and
Astrophysics, University of California, Santa Cruz, California 95064}

\author{Wendy L. Freedman} 
\affil{The Observatories of the Carnegie Institution of Washington,
813 Santa Barbara Street, Pasadena, California 91101}

\author{Barry F. Madore} \affil{NASA/IPAC Extragalactic Database,
Infrared Processing and Analysis Center, California Institute of
Technology, Pasadena, California 91125}

\author{Edward A. Ajhar}
\affil{KPNO, National Optical Astronomy Observatories,\footnotemark[2]
P.O Box 26732, Tucson, Arizona 85726}

\author{William A. Baum} \affil{Department of Astronomy, Box 351580,
University of Washington, Seattle, Washington 98195}

\author{Jon A. Holtzman}
\affil{Astronomy Department, New Mexico State University, Box 30001 /
Dept. 4500, Las Cruces, New Mexico 88003}

\author{C. Roger Lynds}
\affil{Kitt Peak National Observatory, National Optical Astronomy
Observatories,\footnotemark[2] P.O. Box 26732, Tucson, Arizona 85726}

\author{Earl J. O'Neil, Jr.}
\affil{Kitt Peak National Observatory, National Optical Astronomy
Observatories,\footnotemark[2] P.O. Box 26732, Tucson, Arizona 85726}

\author{Peter B. Stetson}
\affil{Dominion Astrophysical Observatory, 5071 W. Saanich Rd., RR5,
Victoria, British Columbia V8X 4M6, Canada} 

\begin{abstract}

We present a \vmi color-magnitude diagram for a region $1'-2'$ from the
center of M32 based on {\it Hubble Space Telescope} WFPC2 images.  The
broad color-luminosity distribution of red giants shows that the
stellar population comprises stars with a wide range in metallicity.
This distribution cannot be explained by a spread in age.  The blue
side of the giant branch rises to $M_I \approx -4.0$ and can be fitted
with isochrones having [Fe/H] $\approx-1.5.$ The red side consists of
a heavily populated and dominant sequence that tops out at $M_I
\approx -3.2,$ and extends beyond $V-I=4.$ This sequence can be fitted
with isochrones with $-0.2 <$ [Fe/H] $< +0.1,$ for ages running from
15~Gyr to 5~Gyr respectively.  We do not find the {\it optically}
bright asymptotic giant branch stars seen in previous ground-based
work and argue that the majority of them were artifacts of crowding.
Our results are consistent with the presence of the {\it
infrared}-luminous giants found in ground-based studies, though their
existence cannot be directly confirmed by our data. The tip of the
metal-poor portion of the giant branch occurs at the luminosity
expected if M32 is at the same distance as M31 but is too sparsely
sampled by this data set to provide a precise distance estimate.  At
fainter magnitudes, the rising giant branch is significantly wider
(FWHM$_{V-I}\sim 0.6$ mag down to $M_I\sim-1.0$) than can be accounted
for by photometric uncertainties, again due to a metallicity spread.
There is little evidence for an extended or even a red horizontal
branch, but we find a strong clump on the giant branch itself, as
expected for the high metallicities inferred from the giant branch.
If the age spread is not extreme, the distribution of metallicities in
M32 is considerably narrower than that of the closed-box model of
chemical evolution, and also appears somewhat narrower than that of
the solar neighborhood.  Overall, the M32 {\it HST} color-magnitude
diagram is consistent with the average luminosity-weighted age of
8.5~Gyr and [Fe/H]$\approx-0.25$ inferred from integrated spectral
indices, extrapolated to the same radius and analyzed with the same
population models.

\end{abstract}


\section{Introduction.}

As the nearest example of an elliptical galaxy, M32 serves as an
important beachhead in our campaign to understand the age and
metallicity mixture of stellar populations in elliptical galaxies.
While M32 may be somewhat unusual given its proximity to M31,
spectroscopically it is identical to other faint ellipticals (Faber
1972; Burstein \etal 1984; Bica, Alloin, \& Schmidt 1990; Gonz\'alez
1993; Trager \etal 1996).  A key objective is to obtain deep
color-magnitude (CM) diagrams to test conclusions thus far based
solely on integrated colors and spectral index work.  The proximity of
M32, combined with the superb resolution afforded by the {\it Hubble
Space Telescope} Wide Field/Planetary Camera 2 (WFPC2) permits a
remarkable improvement in our ability to study its stellar content
and brings us within range of this goal.

In studying a stellar population, the basic variables we wish to
determine are the mean stellar age and metallicity, and the
distribution functions around these
means\footnote[4]{For some elliptical
populations, the Fe-peak/$\alpha$-element ratio also appears to vary
(Worthey Faber, \& Gonz\'ales 1992), but fortunately M32's element
ratios seem close to solar, with Fe/Mg perhaps just slightly
enhanced.}. For old stellar populations like that in M32, it is
difficult to measure both age and Z simultaneously from integrated
light, as broad-band spectral shape and metal lines tend to vary
together in lock step with both these variables (Worthey 1994). To
break this ``age-Z degeneracy", it is necessary to add another
measured quantity that is specifically sensitive to age. Four such
quantities have been suggested for M32: 1) Balmer line strength
(Burstein \etal 1984, Faber \etal 1992), which is directly sensitive
to turnoff temperature, 2) Sr~II$\lambda 4077 /$Fe~I$\lambda 4045$
(Rose 1985, 1994), which is sensitive to the ratio of dwarf to giant
light and hence to turnoff luminosity, 3) ultraviolet
surface-brightness fluctuations at 2500\AA ~(Worthey 1993), again
sensitive to turnoff luminosity, and 4) (1500-V) color (Bressan,
Chiosi, \& Tantalo 1996), which measures age through its effect on the
number and luminosities of hot post-RGB stars.

Many integrated light studies have claimed to have found evidence for
a relatively young, few-Gyr-old stellar population in M32 (\eg Baum
1959, Faber 1972, O'Connell 1980, Burstein
\etal 1984, Rose 1985, Bica, Alloin \& Schmidt 1990, Hardy \etal
1994).  However, in retrospect it is clear that some of these
investigations lacked the basic spectral information needed to
distinguish age and Z, while others lacked sufficiently reliable
population models to correctly interpret the important spectral features.

Rapid progress is being made on both fronts. Stellar population models,
taking both age and Z into account, have been constructed by Worthey
(1994), Bressan \etal (1994), Buzzoni (1995), and by Tantalo \etal
(1996). Accurate measurements of H$\beta$ (Gonz\'ales 1993, hereafter
G93) and Sr/Fe (Rose 1994) in M32 are also now available (the former
as a function of radius out to 1 R$_e$). The upshot of this recent work
is to confirm the earlier findings of relatively young stars, but the
conclusion is now considerably more secure.

Nevertheless, there are important assumptions in these population
models that need to be checked against a real CM diagram.  For
example, Worthey's models for the age and Z of M32 assume a pure red
clump for the horizontal branch (HB). If the H$\beta$ were
concentrated at intermediate F-type temperatures, the need for a young
population would be completely removed (Burstein \etal 1984). All
models also assume a luminosity function for RGB and AGB stars that
crucially affects the predicted red and infrared continua, spectral
features, and surface-brightness fluctuations. Finally, the CM diagram
permits measurement of the spreads about mean properties in a way that
integrated light can never do. These spreads are as important as the
mean values in decoding the star-formation history of the galaxy.

This paper uses a new upper CM diagram of M32 from {\it HST} to address
these and other fundamental aspects of the stellar population. The
major focus is on the metallicity and its distribution.
That is because giant-branch loci are much more sensitive to Z than
age, so conclusions about Z are relatively independent of the
assumed age. The breadth of the giant branch implies a {\it wide}
range in Z, from roughly solar down to below $-1.0$ dex. Paradoxically
though, the distribution is rather narrow when judged by its FWHM
compared to either the closed-box model of chemical enrichment or the
distribution of metallicity in the solar neighborhood.  Evidence for a
wide giant branch and broad total range in Z was available before from
pioneering ground-based CM diagrams in {\it V} and {\it I} by Freedman
(1989) and Davidge \& Jones (1992). However, it is clear now that the old
data were just the tip of the iceberg --- the present data give a {\it
much} fuller picture of the upper giant branch, and a correspondingly
greater total width in \vmi.

A second focus is comparison with mean population age and Z deduced
from ground-based integrated spectral indices. We can infer these
values using Worthey models for the {\it HST} field by slight outward
extrapolation of G93's data. The resulting age-Z pair, 8.5 Gyr and
$-0.25$ dex, agrees perfectly with the color of the giant-branch stars
in the CM diagram.  This indicates a degree of consistency in the
Worthey models but does not independently establish the age. Ages
could be inferred from very deep observations of main sequence turnoff
stars, but crowding precludes this in M32 inside about 3 arcmin, even
with {\it HST}.  Since upper giant-branch tracks are degenerate in age
and Z much like integrated colors and metallic lines, it is not
possible to derive an age from the upper CM diagram alone.  Aside from
verifying the red clump assumption of the Worthey models (in itself an
important contribution), the new CM diagram has little to say about
age.

Nonetheless, the new observations represent an enormous gain over
previous ground-based efforts in our ability to probe the stellar
population of M32. They offer another tantalizing glimpse of
how this so very interesting galaxy may have formed, and 
point the way for future work.

\section{Observations. \label{sec:obs} }

Images of M32 were taken on October 22, 1994, with WFPC2.  Two fields
were imaged, one of M32 (POS1) and the other of a ``background" field
(POS2) to determine the colors and magnitudes of the M31 stars that
make up the majority of field stars at POS1. The POS1 coordinates were
chosen so that the nucleus of M32 was roughly centered in WF3.  The
Planetary Camera (PC) image (0\Sec0455 per pixel) covers a region
extending from 1 to 2 arcminutes almost due south of the center of M32
(see Figure \ref{fig:plate1} = Plate 1).  The POS2 field is situated
$\approx 5$\Min6 NE of POS1 in an area originally intended to match
the background surface brightness of M31 underlying
M32\footnote[5]{Due to a pointing error, this match was not quite
achieved --- see below.}.  Four 500s exposures were taken in each of
F555W ({\it V}) and F814W ({\it I}) in each pointing. An additional
three 1000s exposures were taken through F300W, but we defer
discussion of these data to a later paper.  The coadded F814W image is
shown in Figure \ref{fig:plate2} (Plate 2).


Stellar photometry was carried out using the crowded-field,
PSF-fitting package ALLFRAME (Stetson 1994), the most recent
development in the DAOPHOT series of photometry packages.  ALLFRAME
differs from its predecessors primarily in its ability to use
information from many individual frames simultaneously. The extra
geometric and photometric information available from multiple frames
extends the range of magnitudes and crowding conditions for which
useful photometry is obtainable.  ALLFRAME is applied to the data {\it
after} object detection and aperture photometry are carried out using
DAOPHOT II (Stetson 1987).  The global characteristics of the
CM diagram produced using DAOPHOT II and ALLSTAR were
essentially identical to those found using ALLFRAME. However, the
latter produced fewer outliers and gave a cleaner-looking result.

To push the star detection limit to as faint a level as possible, we
coadded all images taken through the same filter.  ALLFRAME was then
applied to the F555W and F814W images simultaneously. While analyzing
the coadded images does not make full use of the information available
in all 8 individual frames, experiments using both methods showed that
our completeness fraction was significantly higher when coadding
frames prior to object detection.

Neither POS1 nor POS2 had a sufficient number of bright and reasonably
isolated stars to allow proper characterization of the PSF. The PSF
was consequently derived from F555W and F814W images of the globular
cluster NGC 6397 observed with {\it HST} approximately two months
prior to the date of the M32 observations.  No focus changes were
commanded during the intervening period, and desorption of volatiles
will not have significantly altered the focus over this time.

Owing to severe crowding, we ran three DAOPHOT II detection passes to
identify and measure faint stars and to improve the photometry for the
bright stars. After each pass, we subtracted all detected stars from
the images and reran the finding algorithm to search for any
newly-revealed faint stars.  Three such passes yielded a candidate
list of roughly 24,000 objects. Passing this list on to ALLFRAME
resulted in the elimination of $\approx 4000$ objects, leaving a total
of 19,969 measured stars. The final CM diagram is
approximately one magnitude deeper at the 50\% completeness level than
it had been using a single detection pass.

Identical procedures were used to find and measure stars in the POS2
field, yielding a total of 4111 stars.  Under ideal conditions,
correcting for M31 background stars in the M32 field would require
that, for each star found in POS2, we subtract from the POS1 sample
the star most similar in color and magnitude. However, image crowding
is clearly more important for POS1 than POS2, and proper accounting
needs to be taken of their differing levels of completeness. Moreover,
owing to a miscalculation, the POS2 field as observed lies about 57''
outwards from the M31 isophote that passes through POS1. The number of
M31 stars we see in POS2 consequently underestimates the number we
would expect to find in POS1. From interpolation of blue M31 isophotes
plotted by Hodge \& Kennicutt (1982) and by deVaucouleurs (1958) and
from color-index data by Walterbos \& Kennicutt (1988), we estimate
the {\it V} surface brightness gradient of M31 in the vicinity of M32
to be approximately 0.0035 mag/arcsec. From this we conclude that the
stellar surface density expected at POS1 is 1.2 times that found at POS2.

Completeness tests were carried out by adding $\approx 500$ artificial
stars, with colors and magnitudes selected from grid points in the CM
diagram, to each of the POS1 F555W and F814W images. The pixel
locations of the added stars were randomly chosen and identical
between F555W and F814W frames. The frames were then processed using
DAOPHOT II and ALLFRAME in a manner identical to that applied to the
original data.  The results of these tests are shown in Figures
\ref{fig:compfig} and \ref{fig:returned}, which show that the 50\%
completeness level for most POS1 stars occurs at $I\approx 24.9$ and
$V\approx25.4.$ Virtually identical tests were carried out for POS2.

The M32 CM diagram was corrected for contaminating M31 stars as
follows. For a star at POS2 of magnitude {\it I} and color \vmi, a
number of stars

$$N = 1.2 \times C_1(I,V - I)/C_2(I,V - I)$$

\noindent of similar magnitude and color were subtracted from the POS1
sample, where $C_1$ and $C_2$ denote completeness fractions computed
at POS1 and POS2, respectively, and the factor 1.2 corrects for the
misplacement of POS2. In most cases $N$ is not an integer, and stars
were subtracted once, twice, or not at all as determined by whether a
randomly generated number fell above or below the fractional portion
of {\it N}. In total, about 3600 stars were subtracted from the POS1
sample to account for contamination by M31.

The completeness tests also showed that the level of crowding leads to
relatively large photometric uncertainties, ranging from $\pm0.19$ mag
RMS at $I = 22.0$ to $\pm0.46$ mag RMS at $I = 25.0.$ However, the
uncertainties in {\it V} and {\it I} are correlated, and the
color uncertainties range from $\pm0.08$ mag RMS at $I = 22.0$ to
$\pm0.32$ mag RMS at $I = 25.0.$ This is illustrated in Figure
\ref{fig:returned}, where we have plotted the colors and magnitudes of
artificial stars as returned by the DAOPHOT/ALLFRAME processing
sequence.

\section{Discussion \label{sec:discussion}}

\subsection{Morphology of the Color-Magnitude Diagram}

The CM diagrams for the POS1 and POS2 fields are shown in Figures
\ref{fig:cm} and \ref{fig:m31cm}.  For stars bluer than \vmi = 3 the
magnitudes have been transformed from F555W and F814W to
Johnson-Cousins {\it V} and {\it I} using the coefficients of Holtzman
\etal (1995). The \vmi color terms determined by Holtzman \etal are
poorly constrained for {\it very} red stars. They are approximately
zero at \vmi = 2.5 (the reddest color for which good measurements
exist) and we have consequently adopted zero color terms for all stars
with \vmi$ > 3.0$.  The magnitudes have also been dereddened assuming
E({\it B-V}) = 0.08 (Burstein \& Heiles 1982) and using the
absorptions tabulated by Holtzman \etal (1995) for stars of K5
spectral type. An electronic version of the photometry table is
available on request from CJG.

The most striking feature in Figure \ref{fig:cm} is the obviously
composite nature of the CM diagram --- there is a wide spread in the
colors of the giants.  Particularly interesting is the strongly
enhanced red sequence that bends over at $I\approx 21.2,$ and extends
to extremely red ($V-I> 4$) colors.  While Freedman (1989) and Davidge
\& Jones (1992) showed that the giant branch covers a wide range in
color, this extended sequence was not visible in either previous work
and represents a significant change in the picture of the M32 RGB.
The present CM diagram is consistent, however, with Freedman's noted
incompleteness at faint {\it V} magnitudes and her corresponding
insensitivity to very red stars.  This insensitivity is obvious when
one considers the effect of a {\it V}-magnitude limit given the
morphology apparent in the {\it V} vs \vmi diagram, illustrated in
Figure \ref{fig:cmtracks}.  In short, the new {\it HST} measurements
reveal the full extent and morphology of the upper giant branch for
the first time.  Even with {\it HST,} however, there may be some
incompleteness for very red stars with \vmi $ > 3$.  For example, we
find 14 starlike objects brighter than $I=22$ that have no visible
counterparts at all in the F555W image.  Most of these are probably
extremely red giants with \vmi$>4$).

Below about $I = 22,$ the giant stars appear to merge into a rather
fat giant branch of more or less uniform width down almost to the
clump. Based on the completeness tests, the width of the giant branch
above {\it I} $\sim 24$ is significantly larger than can be accounted
for by photometric errors (the color uncertainty at $I = 23$ is 0.125
mag RMS, whereas the observed giant branch width at this point is
0.253 mag RMS --- we return to this point below).  The core-helium
burning stars seem to be concentrated in a red clump.  We see no
indications of RR Lyraes or an extended reddish horizontal branch,
even though the completeness in the region of the diagram where they
should appear ({\it I} $\sim 24.6$, \vmi $\sim 0.5$) is $\sim$65\%.
The form of the fall-off in the density of stars below $I = 25$ is due
to a combination of completeness effects and photometric
uncertainties.

\subsection{Crowding and the Brightest Giants \label{sec:crowding}}

We observe a significant number of stars to be brighter than the first
ascent RGB tip. These may be AGB stars.  However, their luminosities
are significantly fainter than the AGB stars identified by Freedman
(1989).  Assuming a distance modulus $(m-M)_0 = 24.43$ (Ajhar \etal
1996), we expect the tip of the giant branch for older/more metal-poor
populations to occur at $I\approx 20.4$.  We find 23 stars with $I <
20.4$ in POS1 and only one such star in POS2, giving a surface density
of AGB stars\footnote[6]{We use the term AGB loosely here, meaning all
stars brighter than the {\it metal-poor} RGB tip.  Metal rich AGB
stars may be fainter than {\it I} = 20.4, so the surface densities we
derive will necessarily underestimate the true number of
post-horizontal branch stars at all metallicities.} of 70
arcmin$^{-2}$.  Though the absence of such stars in POS2 argues that
these stars must be associated with M32, they do not appear to be as
strongly concentrated towards the center of M32 as the integrated
light.  Freedman finds 91 stars brighter than $I = 20.4$ in her inner
field (a 1.42 arcmin$^2$ region of her field which encompasses POS1
--- see Figure \ref{fig:plate1}), giving a very similar surface
density of 64 arcmin$^{-2}$.

However, whereas our {\it I}-band luminosity function goes essentially
to zero at $I = 19.8,$ both Freedman (1989) and Davidge \& Nieto
(1992) see significant numbers of stars {\it brighter} than $I = 19.8$
extending to $I\approx18$.  Since Freedman's inner field overlaps with
our own, we can test directly the hypothesis that this inconsistency
is due to differences in resolution.  We therefore convolved our PC
image with a Gaussian kernel of FWHM= 0\Sec6 to match Freedman's CHFT
resolution.  We then integer-binned the data to a scale of 0\Sec18
pix$^{-1}$, reasonably close to the 0\Sec2 pix$^{-1}$ scale at the
prime focus of the CFHT.  Finally, we scaled the counts to match the
exposure times used by Freedman, adding $\sim3$ ``background" counts
per pixel as well as noise to match the gain and readout noise
characteristics of the RCA chip on the CFHT. To characterize the
degraded PSF in the standard way, the same process was applied to the
NGC 6397 observations, constructing a PSF from 27 stars still visible
in the resulting image. We applied two passes of the DAOPHOT II
star-detection routine and used ALLSTAR to carry out the PSF-fitting
photometry. The results yielded 147 stars detected in both the F555W
and F814W frames in the region of overlap between the PC field of view
and that of Freedman.  Of these, 68 could be matched with the
photometry of Freedman, yielding a mean {\it I}-band magnitude
difference of only $0.03 \pm 0.35$~mag. The fact that we could match
only half the stars in the field we attribute to differing
completeness levels, which depend on rather small differences in the
noise properties.

The resulting CM diagram and luminosity function are
remarkably similar to those obtained by Freedman.  While the
photometry of the original {\it HST} image yields only 3 stars brighter
than $I = 20,$ smoothing the image results in a total of 19 stars
measured to be brighter than this magnitude. Moreover, the mean color
measured in the degraded image of the stars brighter than $I = 22$
becomes bluer by 0.5 magnitudes, from $V-I = 2.3$ to $V-I = 1.8.$
This strongly suggests that most of the bluer bright giants appearing in
the ground-based CM diagrams of this region are in
fact blends of fainter stars.

The existence of the {\it infrared}-luminous AGB stars found by
Freedman (1992) and Elston \& Silva (1992) using {\it J} and {\it K}
photometry appears to be compatible with our observations, although we
cannot identify them uniquely with $V$ and $I$ alone.  Freedman finds
125 stars with $K<18$ in a $100''\times40''$ field at similar distance
from the nucleus as our {\it HST} observations; 85 of these $K$-bright
stars are also detected in $J.$ Given our smaller image area, we would
expect to have $\sim35$ stars with $K<18$ in our PC observations, with
half of these having $K<17.$ Most of these stars have $J-K>1.1,$ and
would thus be expected to have $I-K>3.5$ if the Milky Way bulge giants
of Frogel \& Whitford (1987) can be used as a guide.  We would
conclude that most of the $K$-bright stars thus have $I>20,$ placing
them comfortably below the bright tip of our $I$-band luminosity
function.  The Frogel \& Whitford giants further show little
correlation between $V-I$ and $I-K$ for $I-K>3.5,$ so the $K$-bright
stars could fall anywhere within the broad $V-I$ spread at the top of
the $I$-band CM diagram.  In contrast, the brightest and bluest stars
in the Freedman sample might be expected to fall {\it above} the tip
of the present $I$-band luminosity function; however, there are only 8
stars with $K<17$ and $J-K<1.1$ in the Freedman sample, giving an
expectation of just 2 such stars for the PC field.  The interesting
problem posed by the Freedman (1992) and Elston \& Silva (1992) data is
that the majority of their $K$-bright stars actually have $J-K$ colors
much redder than the Frogel \& Whitford (1987) bulge sequence, making
them presumably extremely red in $I-K$, and perhaps also $V-I.$ As noted
above, there are 14 stars with $I<22$ that have no $V$ counterparts
--- perhaps these are stars that correspond to a significant fraction
of the $K$-bright sample, and for which it will be very interesting to
obtain NICMOS observations.

\subsection{The Metallicity Distribution}

We argue that the composite nature of the giant branch in M32 is due
to a wide range in metallicity.  Gradients in the spectral index
observations (see below) imply that the (light-weighted) mean age of
the M32 population increases with radius, so there may well be a
mixture of ages present in the PC field; however, a spread in age can
account for only a modest amount of the spread in giant properties
implied by the CM diagram morphology, as we now show.  In Figure
\ref{fig:cmtracks}, we show a set of three isochrones (Worthey 1994)
with ages 2, 5, and 15 Gyr.  Since the three tracks can be made to
overlap by making only modest metallicity changes among them, the age
structure is impossible to deduce from the color distribution of the
giant branch alone.  The tradeoff between age and metallicity for any
assumed RGB track is approximately $d\log|\tau|=-1.7d\log|Z|,$ where
the minus sign indicates that a younger age is counteracted by
increased metallicity.\footnote[7]{\small Note that the behavior here
is close the the behavior \(d\log|\tau| = -1.5 d\log|Z|\) found for
integrated colors and metal-line strengths (Worthey 1994.)} On the
other hand, since age effects are less important than Z, the color
spread must primarily reflect the intrinsic metallicity distribution,
with ambiguity in age causing only minor uncertainty.  This is evident
in Figure \ref{fig:cmtracks}, where we show RGB tracks for clusters
with diverse metallicities, and in Figure \ref{fig:cmtheory}, where we
show theoretical RGB tracks covering a wide range in both metallicity
and age.

We can demonstrate the degree to which metallicity must predominate
over age effects as follows. At $M_I$ = -1.6, the width of the red
giant branch is FWHM(\vmi) $\approx$ 0.6 mag, giving $\sigma$(\vmi) =
0.26 mag. The photometric uncertainty at this magnitude is
$\sigma$(\vmi) = 0.14 mag, yielding an intrinsic RGB width of
$\sigma$(\vmi) = 0.22 mag, or FWHM(\vmi) = 0.51 mag. Now from Figure
\ref{fig:cmtheory} we see that FWHM(\vmi) for $\Delta Z = 1.2$ dex is
$\approx 0.5$ mag, giving a change in color with respect to
metallicity of 0.42 mag/dex. From the scaling rule above, the age
dependence therefore goes as 0.42 / 1.7 = 0.25 mag/dex. Assuming that
the observed width must be the quadrature sum of both effects, we
define

$$\Delta_t \equiv {\rm FWHM ~in} ~\log \tau$$

\noindent and

$$\Delta_Z \equiv {\rm FWHM ~in} ~\log Z$$

\noindent Thus we have

$$ [0.42 \Delta_Z]^2 + [0.24 \Delta_\tau]^2 = 0.51^2.$$

\noindent Setting a maximum allowable age range of $\log \tau =$1.0 dex
(1.5 to 15 Gyr) and solving, we find a minimum allowable $\Delta_Z =
1.1$ dex. The data clearly mandate a large range in metallicity
irrespective of the range in stellar ages which may be present.

The CM diagram implies that the distribution of
metallicity is fairly smooth, but with many more metal-rich stars than
metal-poor ones.  This is illustrated in Figure \ref{fig:hist}, where
we compute the metallicity histogram by laying down a mesh of 10~Gyr
isochrones on the CM diagram covering a wide range of [Fe/H], and then
counting the stars between them brighter than $M_I=-2.4$, where the
uncertainties are fairly small. A small correction for differences in
RGB lifetime at different metallicities was also applied.  No
correction was applied for {\it V}-band incompleteness for the reddest
giants, so the highest metallicity bins slightly underestimate the
true number of giants by up to 14 stars.  One sees that the majority
of stars lie along the red side of the RGB, with a decreasing tail
toward blue, metal-poor RGB loci.  There are very few stars ($<1$\%) more
metal-poor than [Fe/H]$\approx -1.5$.  Although some of us initially
had a visual impression of bimodality in the CM diagram, with a slight
paucity of stars between the main ``blue'' and ``red'' RGBs, the
[Fe/H] distribution shows no evidence of this.  The illusion probably
stems from the fact that RGB color begins to change rapidly with
metallicity at about [Fe/H]$ =-0.5$ and higher.  The effect of
assuming a population age of 15 Gyr is also shown; the histogram
shifts only slightly to lower metallicity.

In Figure \ref{fig:hist} we compare the distribution of metal
abundance found in M32 with closed-box models of chemical evolution.
These simple, one-parameter models are one-zone, with no gas infall or
outflow and zero metal content initially, and assume instantaneous
recycling of heavy elements (\eg Searle \& Sargent 1972).  The
photometric uncertainties for the stars considered here are relatively
small, and the models have not been smoothed. Quite striking is the
fact that the observed metallicity distribution is significantly
narrower than the closed-box models would predict. This region of M32
appears at first glance to be more monometallic than even the solar
cylinder.  On the other hand, if the main ridge of stars in the CM
diagram is comprised of populations with a spread in age, then the
youthful stars may move to the next higher [Fe/H] bin, and it is
possible that the real abundance distribution resembles that of the
solar neighborhood more than it first appears in Figure
\ref{fig:hist}.  For example, the metal abundance of the main red
component is [Fe/H]$= -0.2$ if one assumes an age of 15~Gyr,
[Fe/H]$=-0.07$ for 8~Gyr, but [Fe/H]$=0.01$ for 5~Gyr.  Even so, the
general shape of the distribution remains more or less constant with
age, and the result that there are few low metallicity stars is
secure.

A similar metallicity distribution derived for the POS2 M31 background
field is also illustrated in Figure \ref{fig:hist}, renormalized by a
factor of 5.7 so that the total number of stars is the same as that
for the corrected POS1.  Relative to M32, the M31 population has a
peak at similar Z but a higher tail towards lower metallicities. The
POS2 M31 diagram resembles the solar neighborhood more closely, with
the caveat that the age structure is completely unknown.  The paucity
of metal-poor stars relative to the closed-box model is also inferred
for the nuclei of M31 and M32 (Worthey, Dorman, \& Jones 1996), so it
is probable that this condition exists at most radii in both M32 and
M31.

Worthey et al. (1996) argue that the simplest explanation for the lack
of low-mass, metal-poor stars is that normal processes of chemical
enrichment typically operate to produce fewer low-metallicity stars
than the simplest closed-box model predicts.  That is, modifications
to the simplest model (like variable yield, spatially inhomogeneous
enrichment, or variable IMF schemes that result in fewer metal poor
stars) may be required to mimic chemical enrichment in the real
universe.  However, it is also possible that metal-poor stars could be
present at larger radii.  Large-radius storage of metal poor stars is
a prediction of models of monolithic galaxy collapse (e.g. Larson
1975, Matteucci \& Tornamb\`e 1987, Arimoto \& Yoshii 1987) in which
the metal poor stars are created at large radius and enrichment
proceeds at smaller and smaller radii until most of the gas is
consumed. The abundance distribution at a given radius predicted by
these models is always narrower than the closed-box model (Larson
1975).

The predicted number of metal-poor stars in the collapse model can be
expressed as a fraction of total galaxy mass. In M32, the radius of
the POS1 field encloses about 75\% of the total light, using the
parameters of Kent (1987) and assuming a deVaucouleurs $r^{1/4}$
profile.  Conversion to mass requires a $M/L$ ratio as a function of
radius.  Older ages and higher metallicities drive $M/L$ up, so most
galaxies are inferred to have a higher $M/L$ in the nucleus.  However,
M32 is likely to be an exception to this general rule owing to the
sizeable subpopulation of young stars resident in the nucleus (see
Section \ref{sec-int}).  Various population mixtures were modeled and
compared to both the nucleus and the POS1 positions, and the
best-fitting $M/L_B$ ratios were roughly $M/L_B=3$ to 4.5 on a scale
where Galactic globulars have $M/L_B=2.7$ for both M32 locations
depending on the age-metallicity mixture.  The amount of mass enclosed
within the radius of POS1 (which we have taken to be 75\%) is
consequently about 10\% more uncertain than our estimate of the
enclosed light.

In the closed-box model, about 10\% of the stars in M32 should be more
metal-poor than [Fe/H]$=-1$ assuming an approximately solar yield (the
percentage would be higher if the yield is less than solar). Thus, if
we stipulate a largely circular orbit distribution for metal-poor
stars to account for the lack of such stars in our sample, there would
still appear to be sufficient room to store them in the outer $\sim
25$\% of the galaxy at $R > 2^\prime$. {\it HST} observations are
scheduled for a field in the outskirts of M32 that encloses about 92\%
of the light.  If these observations uncover copious stars of
[Fe/H]$=-2.5$ to $-1.5,$ this would strongly support an outside-in
dissipational origin for M32. If the observations find few such stars,
then either they did not form in large numbers, as Worthey et al.
suggest, or they were tidally stripped by close passages with M31, a
possibility not without supporting evidence (Faber 1973; Nieto \&
Prugniel 1987).

\subsection{Clump Morphology}

Turning back to the CM diagram, the morphology of the core-helium
burning stars in M32 is a ``clump'' instead of a ``blue,''
``extended,'' or ``red'' horizontal branch as commonly seen in
Galactic (and M31) globular clusters with metallicity less than around
$-0.7$. This confirms the conclusion of Rose (1994) that a clump,
rather than red horizontal branch stars, must dominate the core-helium
burning stars to produce the observed spectral line indices and
broad-band colors.  A ``clump'' morphology is expected for metal-rich
populations (Hatzidimitriou 1991), so the fact that the M32 clump is
strong is consistent with the giant branch color distribution.
 
The extent to which the core-helium stars lie exclusively within a
clump can in principle set upper limits on the numbers and ages of
stars with metallicities between [Fe/H]$= -0.8$ and $-1.5.$
Superficially, the CM diagram is consistent with a pure-clump
population.  If {\it all} core-helium burning stars are in the clump,
then the metal-poor stars in M32 must be younger than $\sim$10~Gyr, as
populations this young or younger are neither observed (Stryker, Da
Costa, \& Mould 1985) nor predicted (e.g. Lee, Demarque, \& Zinn 1994)
to have RR Lyrae stars or stars intermediate in temperature between
the instability strip and the clump.  This picture is contrasted
against one in which most of the stars in M32 are very old. In that
case the more metal-poor components should display bluer horizontal
branches like the ``red'' one of 47 Tuc or the ``extended'' one of M3
(a theoretical approximation of which is shown in Figure
\ref{fig:cmtracks}).
 
Unfortunately, there are so few stars with [Fe/H]$<-0.7$, and the
photometric errors are so large at the base of the CM diagram, that it
is not possible to detect or rule out blue horizontal branch stars
unambiguously.  Using Figure \ref{fig:hist} as a guide, and assuming a
15 Gyr population, there are 128 stars more metal-poor than
[Fe/H]$=-0.7$ (out of 842 brighter than $M_I=-2.4$).  The ratio of the
number of bright giants corresponding to the Figure \ref{fig:hist}
cutoff to the number of horizontal branch stars in either M3 or 47 Tuc
is about 0.5, so we would expect over 200 metal poor horizontal branch
stars to be present under the all-15-Gyr hypothesis.  To test whether
we see these stars or not, false stars were added to the observed CM
diagram in either a 47-Tuc morphology or an M3 morphology, with random
errors approximately as observed.  The conclusion from this exercise
is that up to several hundred 47-Tuc-like HB stars could exist in the CM
diagram, yet be hidden in the broad envelope of the more numerous true
clump stars.  Bluer M3-like HB stars could also exist, but probably
less than 200 of them.  In short, the present CM diagram cannot
strongly constrain the upper age limits of the most metal-poor stars
in M32.

\subsection{Age Information from Integrated Light} \label{sec-int}

Although the present CM diagram yields a clear picture of the
metallicity distribution in M32, it does less well regarding age,
especially of any younger components that might be present.
Integrated colors provide some information on age if the metallicity
distribution is approximately known, but the answer is intrinsically
slippery because of model-to-model differences in the prediction of
colors, which translate into a 35\% scatter in age estimates (Charlot,
Worthey, \& Bressan 1996). The intrinsic power of the isochrones to
delimit age is also not that much higher than integrated colors:
$\Delta \log Z = -1.7 \Delta \log \tau$ (see above) vs. $\Delta \log Z
= -1.5 \Delta \log \tau$ (Worthey 1994). In addition, with integrated
indices (including colors), we are presently limited to deriving an
average age rather than a star formation history.
 
Worthey (1994) has demonstrated that, to break the age-Z degeneracy,
it is necessary to add information beyond colors and metal lines, for
example, Balmer line strength. Using this approach, we will first
attempt to derive an age under the simplifying assumption that the
population is characterized by a single age. G93 took long-slit
spectra to obtain line-strength indices on the Lick system, which can
be directly compared to the Worthey (1994) model predictions. Over the
POS1 PC field of view, the G93 gradients (extrapolated from
$45^{\prime\prime}$) were averaged, weighted by amount of light:

$$<I>={{\int_{\rm area} I \mu_r{\rm d}A }\over{\int_{\rm
area}\mu_r{\rm d}A}}$$ 

\noindent where $<I>$ is the average index, $\mu_r$ is the $r$-band
surface brightness as a function of position from Kent (1987), and $A$
is the area over which the averaging was performed.  Given the shallow
color gradients in M32, $r$-band will adequately mimic $V$, where the
indices are actually measured.  The light-weighted mean radius of the
PC field is 1.8 $R_e$, where G93 H$\beta =1.92,$ Mg b =2.99, and
$<{\rm Fe}> =2.42.$ Worthey (1994) models predict that H$\beta$
relative to the combined index [MgFe] = (Mg b $\times<{\rm
Fe}>)^{1/2}$ can give a simultaneous estimate for a mean age and a
mean metallicity, with an age in this case of 8.5 Gyr and
[Fe/H]$=-0.25$ for $R = 1.8 R_e$.

This is noteworthy because if 8-Gyr isochrones are laid down on the
POS1 CM diagram and a metallicity distribution computed as in Figure
\ref{fig:hist}, the mean metallicity of the stars is [Fe/H]$=-0.25$.
This demonstrates (at least) internal model coherence as shown by the
close agreement in mean metallicity estimates from star counts in the
CM diagram on the one hand vs. line strengths on the other.  Within
model errors, which can be substantial (Charlot, Worthey, \& Bressan
1996), the integrated colors of the M32 POS1 field are also consistent
with the same age and Z.

Toward the nucleus of M32 the metallic indices of G93 get only
slightly stronger, while H$\beta$ increases in strength dramatically
(confirmed by Hardy et al. 1994). Differentially within the models,
the only way H$\beta$ {\em and} metallic indices can both increase in
strength is if the mean population towards the nucleus is
simultaneously younger and more metal rich. The G93 indices at the
nucleus indicate a mean age of $\sim$4 Gyr and a metallicity just less
than solar. To within the uncertainties, {\em the predicted color at
the nucleus is the same as that at 1.8 $R_e$,} despite a factor of two
difference in predicted age because the age effect is balanced by an
average abundance enhancement.

Although age is the simplest explanation for the observed H$\beta$
gradient and the flat metallic feature strength and color gradients,
two other alternatives should be considered: large numbers of either
blue straggler stars or blue HB stars, if concentrated toward the
center, could increase H$\beta$ strength as observed. However, both of
these alternatives would make for bluer colors and weaker metallic
features toward the center, which are not observed. Lacking a
plausible mechanism to counter these effects, one would still require
an age or metallicity gradient in M32. In the case of blue HB stars or
young A-type main sequence stars, significant numbers of them appear
to be ruled out in M32's nucleus by the Ca II index (Rose 1994;
Worthey, Dorman, \& Jones 1996). Moreover, in contrast to M32, other
elliptical galaxies generally show strongly increasing metal line
strengths, reddening colors, and slightly weaker or constant H$\beta$
toward the center.  If blue stragglers or blue HB stars are
responsible for M32's high central H$\beta$, we must then explain why
M32 behaves differently than other ellipticals.

Despite these favorable checks, there remains little detailed
information on the age structure in M32 because all the details are
averaged into one or two indicators.  Dropping the assumption of a
single age for the POS1 field, we constructed limited two-age models
to try to quantify the uncertainty.  The models begin with an assumed
base age.  Based on this age, a metallicity histogram is compiled from
the CM diagram data.  Then different age/metallicity components are
substituted for the most populous bin in various proportions, and the
results are compared with G93 indices.  The constructed models are at
all times consistent with the CM diagram for bright giants.

The (merely illustrative, not definitive) result for a base age of 15
Gyr is that about 7\% (by total mass) of a 2-Gyr population is needed
to match the G93 indices. Interestingly, this is consistent with the
results of a near-infrared survey by Silva \& Bothun (1996) of
ellipticals with strong nuclear H$\beta$ and blue central colors. For
a base age of 10 Gyr, 5\% of a 5-Gyr population would suffice.  For a
base age of 8 Gyr, the best match is obtained if there is 25\% to 30\%
of an old, 15 Gyr population present.  From the way these populations
balance, it is clear that {\em over half of the stars at the POS1
radius must be as old or older than 8 Gyr.} All of these models match
the data as well as the single-age 8.5 Gyr model. The conclusion is
that there are many ways of balancing subcomponents so that the
see-saw lands at a mean age of 8.5 Gyr. This sort of modeling, which takes
into account CM diagram information as well as integrated light
indices, is in its infancy.

\subsection{The Tip of the Red Giant Branch}

The tip of the red giant branch (TRGB) in metal-poor populations can
be used as a distance indicator.  Old, metal-poor giant stars have
peak absolute {\it I}-band magnitudes of M$_I\sim-4$ mag.  Da Costa \&
Armandroff (1990) calibrated the magnitude of the TRGB for the range
of metallicities ($-2.2 <$[Fe/H]$< -0.7$) defined by Galactic globular
clusters.  With this calibration, Lee, Freedman \& Madore (1993)
compared TRGB distances to 10 nearby galaxies to those from Cepheids
and/or RR Lyrae stars; they found excellent correspondence at a level
of $\pm10$\% RMS.  Subsequent studies ({\it e.g.,} Sakai, Madore, \&
Freedman 1996a,b) confirm the excellent agreement between the Cepheid
distances and the TRGB distances for metal-poor giants.  A key
property of the TRGB distance indicator is that M$_I$ varies by less
than $\sim0.1$ mag for [Fe/H]$< -0.7.$

Use of the TRGB distance indicator for M32 is problematic, given the
high average metallicity of its population.  To date, no empirical
calibration for the TRGB exists for populations with metallicities
higher than defined by the Galactic globular clusters.  Based on the
models of Worthey (1994), the expected behavior of the TRGB for more
metal-rich systems is shown in Figure \ref{fig:trgb}.  Lee, Freedman
\& Madore (1993) conclude that the method could be applied to systems
{\it as long as the galaxies show an appreciable population of
low-mass, resolved red giant branch stars with [Fe/H] $< -0.7$ dex.}
Figures \ref{fig:cm}, \ref{fig:trgb}, and \ref{fig:lf} underscore the
importance of obtaining color information before blindly applying the
TRGB method.  The redder, metal-rich giants in M32 are almost a
magnitude fainter than the bluer, metal-poor giants, and an uncritical
application of this method to the metal-rich component could lead to a
serious systematic error.  The TRGB method works by passing an
edge-detecting or ``Sobel'' filter over the {\it I}-band luminosity
function that generates a strong signal at locations where the
luminosity function has sharp discontinuities.  As can be seen in
Figure \ref{fig:lf}, the M32 luminosity function over all
metallicities lacks a well-defined edge.  Selecting just the metal
poor stars (those with $V-I<1.8$ mag) does produce a luminosity
function with a sharper edge, but unfortunately in this case, there
are too few such stars in the PC field to define the metal-poor TRGB
accurately.  Application of the edge detector to just the metal-poor
giants yields a TRGB magnitude of m$_I=20.75,$ or M$_I=-3.7$ mag.
This value has a large uncertainty, however, and its difference from
the expected M$_I=-4.0$ (for the assumed distance modulus of 24.43) is
not significant.

\subsection{Luminosity Function}

The {\it I}-band luminosity function (LF) we derive for M32 is shown
in Figure \ref{fig:lf}, along with theoretical LFs for selected
age/metallicity pairs. The model LFs were generated using a pure
clump, as opposed to a red or blue horizontal branch. The mismatch
between the data and the model LFs in the clump region is likely due
to a combination of $(i)$ ill-determined completeness corrections at
very faint magnitudes, $(ii)$ the asymmetric nature of the photometric
uncertainties in this region, which we have characterized as Gaussian
for simplicity, $(iii)$ uncertainties and a possible spread in the
theoretically predicted clump luminosities, and $(iv)$ the possibility
that severe crowding is causing us to overestimate the luminosities of
very faint stars in a manner analogous to that for AGB stars described
in Section \ref{sec:crowding}. Of the four model LFs shown, the
8~Gyr/[Fe/H]$=-0.07$ model is the best match to the observed
discontinuity and bump in the region $ 20.5 < I < 22.0$. However, the
differences among the models are sufficiently small that a slight
change in the assumed distance to M32 could easily negate this
distinction.

There are fairly substantial discontinuities in the LF at $I\approx
20.5$ and $I\approx 21.$ The latter (which is the more significant) is
clearly a consequence of the young/metal rich giant branch, which is
essentially horizontal in Figure \ref{fig:cm} at this magnitude.
Assuming $(m-M)_0= 24.43,$ the discontinuity at $I\approx 20.5$ is
very close to that expected for old/metal poor stars. Indeed, if
we generate a LF using only metal-poor stars bluer than $V-I = 1.8$
(dashed histogram), the discontinuity is considerably enhanced. This
is the tip of the giant branch used in the TRGB distance method. The
cut again illustrates the importance of ensuring that the sample is
not contaminated by stars more metal rich than [Fe/H]$\approx -0.7$
when applying the TRGB method.  As we noted in the previous section,
the Sobel filter actually locates the edge of the metal-poor TRGB at a
slightly fainter luminosity, but with little significance.

To improve our sampling of the bright end of the luminosity function,
we also analyzed the less-crowded half of chip WF4. Using a reduction
procedure essentially identical to that used for the PC, this yielded
another 20,000 stars (this field is slightly further from the center
of M32). With allowances for the greater degree of undersampling and
differing levels of completeness, the morphology of the
CM diagram and the luminosity function obtained in the
WF4 field are completely consistent with the results from the PC.

\section{Summary and Conclusions}

We have analyzed WFPC2 images of M32. Based on ALLFRAME 
crowded-field photometry we conclude the following:

\begin{itemize}

\item{} The observed spread in color among giant branch stars {\it
requires} a substantial range of metallicity.  Allowing for a large
range in population age will have only a slight effect on the range of
metallicity inferred.

\item{} The metallicity distribution is smooth and strongly skewed
towards metal rich stars.  The peak of the metallicity histogram
occurs at $-0.2<$[Fe/H]$<+0.05$ for assumed mean ages in the range
15~Gyr to 5~Gyr for the stellar population.  A low metallicity tail
extends to [Fe/H]$\approx-1.5.$

\item{} The raw metallicity distribution appears somewhat narrower
than that of the solar cylinder although, if there are strong age
admixtures, the distribution might resemble the local one more
closely. In any case, the M32 distribution is much narrower than the
closed-box model of chemical evolution. It remains an open question
whether the missing metal-poor stars are absent entirely from M32 or
whether they are stored at large radii.

\item{} The M31 abundance distribution in the outer disk sampled by
POS2 resembles the solar neighborhood if the stars there are mostly
older than a few Gyr.

\item{} The width of the giant branch remains significant at
magnitudes well below the tip, also supporting the conclusion that the
population has a large dispersion in metallicity.

\item{} There is a strongly enhanced red clump at the bottom of the
giant branch, consistent with Rose's (1994) analysis based on
broad-band colors and spectral indices, as well as the metallicity
distribution implied by the giant branch morphology.  At the same
time, we cannot rule out a small minority population of extended-red
or even blue horizontal branch stars, which would be expected to be
present if the most metal-poor stars in M32 are older than 10 Gyr.

\item{} We do not see the very luminous blue AGB stars found in
previous ground-based investigations.  Experiments reveal that many of
these stars are artifacts of image crowding.  In contrast, a small
population of the {\it K}-bright AGB candidates identified by Freedman
(1992) and Elston \& Silva (1992) may be present in our sample but
cannot be identified by their corresponding {\it I} and {\it V}
luminosities alone.

\item{} Whereas the metal poor giant branch extends to {\it I}
$\approx -4.0$, the metal rich population reaches only to {\it I}
$\approx -3.2$.  This underscores the importance of obtaining color
information and selecting only stars with [Fe/H] $< -0.7$ before
applying the TRGB method to the determination of distances.

\item{} The M32 CM diagram is consistent with the integrated line
indices of Gonz\'alez (1993), extrapolated to the same radius.
Interpreted with Worthey (1994) models, the indices give a mean age of
8.5 Gyr and a mean abundance of $-0.25$ dex. If 8-Gyr isochrones
are overlaid on the CM diagram, the mean stellar abundance from
counting stars as a function of color is also $-0.25$ dex.

\item{} The Gonz\'alez indices indicate a strong radial gradient in
the mean stellar population with radius, the nucleus being much
younger and somewhat more metal rich than the outer parts studied
here. This is because the metallic indices are nearly flat but
slightly increasing toward the center and the broad-band colors nearly
constant, whereas H$\beta$ increases strongly inwards.

\item{} Two-age models are also consistent with both the CM diagram
and the Gonz\'alez (1993) indices, showing that there are many ways to
mix ages and metallicities to match the observations. Depending on the
age of the bulk of the stars, modest subpopulations of quite youthful
stars are allowed. However at least half the stars {\em must} be at
least as old as 8 Gyr. We are hopeful that future work will allow us
to better constrain the options.

\end{itemize}

\acknowledgments

We thank David Silva for discussions concerning the {\it K}-bright
AGB stars, and Shoko Sakai for discussions concerning detection of the TRGB.
This research was conducted by the WF/PC Investigation Definition
Team, supported in part by NASA Grant No. NAS5-1661. One of us (GW) is
partially supported by NASA funds through grant HF-1066.01-94A from
the Space Telescope Science Institute.

\clearpage

\figcaption{Digitized Sky Survey image of the field containing M32.
The outlines indicate the WFPC2 POS1 and POS2 fields and the
Canada-France-Hawaii telescope CCD frame examined by Freedman (1989).
The field shown in Figure \protect{\ref{fig:plate2}} is indicated by
the heavy white line.  The entire field shown here subtends
$15^{\prime}$ on a side.  \label{fig:plate1}}

\figcaption{Linear stretch of the F814W PC frame. The lower-left
corner of the field is 56 arcseconds from the center of M32.
\label{fig:plate2}}

\figcaption{Completeness fraction as a function of {\it I} and {\it V}
magnitudes sampled at 4 different colors. Filled symbols and solid lines
correspond to the POS1 field, while open symbols and dotted lines
derive from completeness tests in POS2.  \label{fig:compfig}}

\figcaption{Colors and magnitudes of artificial stars returned in the
course of completeness tests. The open circles indicate the grid of
input colors and magnitudes, and the points show the colors and
magnitudes returned after processing by DAOPHOT/ALLFRAME.
\label{fig:returned}}

\figcaption{ALLFRAME color-magnitude distribution of stars in the POS1
(M32) field, corrected for background contamination using the
color-magnitude distribution found in the POS2 field. The magnitudes
and colors shown have been transformed to {\it V} and {\it I} using a slightly
altered transformation (zero color terms for the very reddest stars)
from that given by Holtzman \etal (1995). \label{fig:cm}}

\figcaption{ALLFRAME color-magnitude distribution of stars in the POS2
(background) field. \label{fig:m31cm}}

\figcaption{The corrected color-magnitude diagram of stars in M32 as
in Figure \protect\ref{fig:cm}.  Approximate V-band cutoffs are shown
for these {\it HST} data (at $V\approx 26.0$) and for the Freedman (1989)
data (at $V\approx 23.2$). Giant branch fiducials from Da Costa \&
Armandroff (1990) are shown for Galactic globular clusters M15
([Fe/H]$\approx -2.2$), NGC 6752 ([Fe/H]$\approx -1.5$), and 47 Tuc
([Fe/H]$\approx -0.8$). Also shown is the giant branch of the old open
cluster NGC 6791 ([Fe/H]$\approx +0.2$, age$\approx$8 Gyr from
Garnavich et al. (1994)). The reddest point on this fiducial represents
the reddest giant in the cluster, but it may not correspond to the
helium flash point because the RGB is sparsely populated.
Three Worthey (1994) isochrones are superimposed to illustrate the
degeneracy of age and metallicity in RGB color. The solid line is
(age,[Fe/H])$= (15,-0.30),$ dotted $(5,-0.06),$ dashed $(2,+0.15).$
The predicted location of a metal-poor extended horizontal branch is
shown. The line at $M_I=-2.4$ is the cut above which stars were
counted for the histograms of Fig. \protect\ref{fig:hist}.  The error
bars are from artificial star tests and refer only to color $V-I = 1.34$.
\label{fig:cmtracks}}

\figcaption{Worthey (1994) isochrones are shown on the POS1 data for
[Fe/H] between $-1.2$ and 0.0 in 0.2-dex steps. More metal-rich
isochrones are redder. Isochrones aged 8 Gyr are shown as dotted
lines; isochrones aged 15 Gyr are shown as solid lines. The
hypothetical locations of metal poor, extended horizontal branches are
shown for both ages, the younger one nearly 0.2 mag brighter. Two
examples of how bright AGB stars are predicted to lie in this diagram
are shown extending beyond the 15-Gyr RGB isochrones of [Fe/H]$=-0.6$ and
$-0.4$.  The errors are for color $V-I=1.34$ as in Figure
\protect\ref{fig:cmtracks}.  \label{fig:cmtheory}}

\figcaption{Counts of M32 stars caught between 10~Gyr isochrones of
1~different abundance. Stars brighter than $M_I=-2.4$ have small enough
errors and the color change with abundance is large enough to make
this a reliable method of estimating the abundance distribution of
stars in M32. The raw counts between isochrones spaced at 0.2 dex
intervals is shown as the coarse histogram with the heavy line. A
small correction to compensate for the expected variable number of
giants as a function of metallicity was incorporated. The light-line
histogram assumes 15 Gyr isochrones rather than 10 Gyr isochrones. The
dashed-line histogram shows the distribution for POS2 and 10 Gyr
isochrones, and the counts have been multiplied by 5.7 to normalize
the total number of counts to be same as for POS1.  For comparison,
two closed-box models of yield $\log [Z_{\rm yield}/Z_\odot] = -0.5$ and
$-0.2$ are shown. The M32 distribution is much narrower than the
closed-box models. Two derivations of the abundance distribution in
the solar cylinder are also shown (Rana 1991, Pagel 1989). The M32
histogram shown is even more narrow than the solar cylinder
distribution, but if M32 contains many stars which are substantially
younger than 10~Gyr, they will have higher abundance, and should be
plotted (at most) one bin to the right. The true abundance
distribution may therefore resemble that of the solar neighborhood
quite closely, except for near [Fe/H]$=-0.5$, where M32 definitely has
fewer stars than the solar neighborhood. The M32 sample of stars is 7
times larger than the sample used for the solar neighborhood studies.
\label{fig:hist}}

\figcaption{Absolute {\it I}-magnitude of the tip of the red giant
branch as a function of metallicity, from the models of Worthey
(1994).  \label{fig:trgb}}

\figcaption{Background-corrected {\it I}-band luminosity function of the
POS1 field before (solid histogram) and after (filled circles)
applying completeness corrections. The error bars reflect counting
statistics only. The theoretical luminosity functions are shown after
convolution with the estimated photometric uncertainties and
application of the completeness function. The 15~Gyr models are shown
as solid curves while the 8~Gyr models are shown as dotted curves. The
models have been normalized using the number of stars found with
$22.25<I<22.5.$ The dashed histogram shows the effect of counting
only those stars with $V-I< 1.8$.  \label{fig:lf} }


\clearpage

\begin{references}

\reference{} Ajhar, E. A., Grillmair, C. J.,  Lauer, T. R., Baum, W.
A., Faber, S. M., Holtzman, J. A., Lynds, C. R., \& O'Neil, E. Jr.
1996, \aj, 111, 1110 %

\reference{} Arimoto, N., \& Yoshii, Y. 1987, \aap, 173, 23 %

\reference{} Baum, W. A. 1959, \pasp, 71, 106 %

\reference{} Bica, E., Alloin, D., \& Schmidt, A. A. 1990, \aa, 228,
23 %

\reference{} Bressan, A., Chiosi, C., \& Fagotto, F. 1994, \apjs, 94,
63 %

\reference{} Bressan, A., Chiosi, C., \& Tantalo, R. 1996, \aa, in
press %

\reference{} Burstein, D., \& Heiles, C. 1982, \aj, 87, 1165 %

\reference{} Burstein, D., Faber, S. M., Gaskell, C. M., \& Krumm, N.
1984, \apj, 287, 586 %

\reference{} Buzzoni, A. 1995, \apjs, 98, 69 %

\reference{} Charlot, S., Worthey, G., \& Bressan, A. 1996, \apj, 457,
625 %

\reference{} DaCosta, G. S., \& Armandroff, T. E. 1990, \aj,
100, 162 %

\reference{} Davidge, T. J., \& Nieto, J.-L 1992, \apjl, 391, L13 %

\reference{} Davidge, T. J., \& Jones, J. H. 1992, \aj, 104, 1365 %

\reference{} deVaucouleurs, G. 1958, \apj, 128, 465 %

\reference{} Elston, R., \& Silva, D. R. 1992, \aj, 104, 1360 %

\reference{} Faber, S. M. 1972, \aap, 20, 361 %

\reference{} Faber, S. M. 1973, \apj, 279, 423 %

\reference{} Faber, S. M., Worthey, G., \& Gonz\'alez, J. J. 1992, in
The Stellar Populations of Galaxies: Proceedings of IAU Symposium 149,
ed. B. Barbuy, Kluwer Acedemic Publishers, Dordrecht, p. 255 %

\reference{} Freedman, W. L. 1989, \aj, 98, 1285 %

\reference{} Freedman, W. L., 1992, \aj, 104, 1349 %

\reference{} Frogel, J. A., \& Whitford, A. E. 1987, \apj, 320, 199 %

\reference{} Gonz\'alez, J. 1993, PhD Thesis, Univ. California, Santa
Cruz, G93 %

\reference{} Hardy, E., Couture, J., Couture, C., \& Joncas, G. 1994,
\aj, 107, 195 %

\reference{} Hatzidimitriou, D. 1991, \mnras, 251 545

\reference{} Hodge, P., \&  Kennicutt, R.C. 1982, \aj, 87, 264 %

\reference{} Holtzman, J. A., Burrows, C. J., Casertano, S., Hester,
J. J., Trauger, J. T., Watson, A. M., \& Worthey, G. 1995, \pasp, 107,
1065 %

\reference{} Kent, S. M. 1987, \aj, 94, 306 %

\reference{} Larson, R. B. 1975, \mnras, 173, 671 %

\reference{} Lee, M.~G., Freedman, W.~L. \& Madore, B.~F., 1993,
\apj, 417, 553 %

\reference{} Lee, Y.-W., Demarque, P., Zinn, R. 1994, \apj, 350, 155 %

\reference{} Matteucci, F., \& Tornamb\`e, A. 1987, \aap, 185, 51 %

\reference{} Nieto, J. -L., \& Prugniel, P. 1987, \aap, 186, 30 %

\reference{} O'Connell, R. W. 1980, \apj, 236, 430 %

\reference{} Rose, J. A. 1985, \aj, 90, 1927 %

\reference{} Rose, J. A. 1994, \aj, 107, 206 %

\reference{} Sakai, S., Madore, B. F., \& Freedman 1996a, \aj, in
press %

\reference{} Sakai, S., Madore, B. F., \& Freedman 1996b, \aj, in
preparation %

\reference{} Searle, L., \& Sargent, W. L. W., 1972, \apj, 173, 25 %

\reference{} Silva, D. R., \& Bothun, G. 1996, in preparation %

\reference{} Stetson, P. 1987, \pasp, 99, 191 %

\reference{} Stetson, P. B. 1994, \pasp, 106, 250 %

\reference{} Stryker, L. L., Da Costa, G. S., \& Mould, J. R. 1985,
\apj, 298, 544  %

\reference{} Tantalo, R., Chiosi, C., Bressan, A., \& Fagotto, F.
1996, \aa, in press  %

\reference{} Trager, S. C., Faber, S. M., Burstein, D., \& Worthey,
G. 1996, in preparation %

\reference{} Walterbos, R., \& Kennicutt, R. C. 1988, \aap, 198, 61 %

\reference{} Worthey, G. 1993, \apj, 415, L91 %

\reference{} Worthey, G. 1994, \apjs, 95, 107 %

\reference{} Worthey, G., Faber, S. M., \& Gonz\'alez, J. J., 1992,
\apj, 398, 69 %

\reference{} Worthey, G., Dorman, B., \& Jones, L. A. 1996, \aj, in
press %

\end{references}
\end{document}